\documentclass[aps,preprint,superscriptaddress]{revtex4}
\usepackage{graphicx}
\usepackage{amsfonts,amsmath,amssymb,latexsym}

\begin{document}
\title{Energetic Instability Unjams\\ Sand and Suspension}
\author{Yimin Jiang}\email{yimin.jiang@uni-tuebingen.de}
\affiliation{Theoretische Physik, Universit\"{a}t
T\"{u}bingen, 72076 T\"{u}bingen, Germany, EU}
\affiliation{Physics \& Technology, Central South
University, Changsha 410083, China}
\author{Mario Liu}\email{mliu@uni-tuebingen.de}
\affiliation{Theoretische Physik, Universit\"{a}t
T\"{u}bingen, 72076 T\"{u}bingen, Germany, EU}
\date{\today}

\begin{abstract}
Jamming is a phenomenon occurring in systems as
diverse as traffic, colloidal suspensions and granular
materials. A theory on the reversible elastic
deformation of jammed states is presented. First, an
explicit granular stress-strain relation is derived
that captures many relevant features of sand,
including especially the Coulomb yield surface and a
third-order jamming transition. Then this approach is
generalized, and employed to consider jammed magneto-
and electro-rheological fluids, again producing
results that compare well to experiments and
simulations.
\end{abstract}
\maketitle

We start our study of jamming~\cite{Liu-Nagel} in
granular systems, by deriving an appropriate
stress-strain relation from a simple, postulated
elastic energy. It accounts for the reversible elastic
deformation of granular systems, up to the point of
yield, and reproduces many relevant results from
granular physics and soil
mechanics~\cite{deGennes,Kolymbas}, including the
compliance tensor, {\em Rankine} states, and shear
dilatancy. Moreover, the elastic energy is convex only
below the Coulomb yield condition and becomes unstable
there. As a result, the system escapes from the
strained state and looses shape-rigidity, providing an
explanation why sand unjams. Next, the granular
elastic energy is shown to be a special case of a more
generally valid energy expansion, with respect to the
shear strain. Realizing that this expansion may serve
as the starting point to account for other jammed
systems, we use it to consider colloidal
suspensions~\cite{Trappe}, specifically magneto- and
electro-rheological fluids, which solidify at fields
strong enough~\cite{Halsey,Wen}. Again, an energy
expression is proposed, from which the magnetic,
dielectric and elastic behavior is deduced, especially
the solid-fluid phase diagram.

Our basic understanding of sand is due to Coulomb, who
noted that its most conspicuous property is yield: A
pile of dry sand possesses a critical slope that it
will not exceed. His insightful conclusion is that the
quotient of shear stress over pressure must not exceed
a certain value, $|\sigma _s|/P\leq \mu _f$. Wet sand
can sustain a small shear stress $\sigma _c$ even at
vanishing pressure. It satisfies the {\em
Mohr-Coulomb} condition, $|\sigma _s|\leq \mu
_fP+\sigma _c$, see~\cite{sm1}.

It is standard praxis in soil mechanics to calculate
the stress distribution by taking the stress
$\sigma_{ij}$ as some function of the strain $u_{ij}$.
Unfortunately, the calculated stress distribution
routinely contradicts the Coulomb condition, and yield
must be postulated, {\em ex post facto}, where it is
not satisfied. An improvement of this somewhat brute
method is given by the {\em Rankine} states, $\sigma
_s=\pm P\mu _f$, which should hold close to yield. The
ameliorated calculation is given by accepting the
result of elasticity away from the region of failure,
postulating a Rankine state close to it, and
connecting both smoothly. Clearly, in spite of
ingenious ways to circumvent it, the basic problem is
the lack of a stress-strain relation $u_{ij}(\sigma
_{kl})$, with which a realistic stress distribution
can be calculated.

If we had $u_{ij}(\sigma _{kl})$, the incremental
relation, $\delta u_{ij}=(\partial u_{ij}/\partial
\sigma _{kl})\delta \sigma _{kl}$ $\equiv \lambda
_{ijkl}\delta \sigma _{kl}$, is easily derived. The
elements of the compliance tensor $\lambda _{ijkl}$
can also be obtained from experiments, in which
$\delta u_{ij}$, the strain response to a stress
change $\delta \sigma _{ij}$, is
measured~\cite{Kuwano-Jardine}. Although integrating
the measured $\lambda _{ijkl}$ should in principle
lead to $u_{ij}(\sigma _{kl})$, this is a hard,
backward operation -- made more difficult by the
typical scatter of data, partly from irreversible
plastic deformations. This circumstance has led many
to espouse the view that $\lambda _{ijkl}$ is
history-dependent, that an explicit
$u_{ij}(\sigma_{kl})$ (from which to deduce $\lambda
_{ijkl}$) does not exist. Different elasto-plastic
theories, some exceedingly complex, have been
constructed to account for $\lambda _{ijkl}$,
including both elastic and plastic deformations,
though a universally accepted model is
missing~\cite{Kolymbas}.

Confining our study to reversible elastic
deformations, we derive a stress-strain relation to
account for the listed granular behavior. We start
from the elastic energy
\begin{equation}\label{1}
w=\textstyle\frac 12{\delta}^{0.5}(B\delta^2+Au_s^2),
\end{equation}
where $\delta\equiv -u_{\ell \ell }$ is the
compression, $u_s^2\equiv u_{ij}^0u_{ij}^0$ is shear
strain squared. ($u_{\ell \ell }$ denotes the trace of
the strain and $u_{ij}^0$ its traceless part. $\delta,
u_s=0$ imply the grains are in contact but not
compressed or sheared.) $A, B>0$ are functions of the
void ratio $e$, an independent variable. We adopt the
same empirical expression for both, $A,B\sim
(2.17-e)^2/(1+e)$, see~\cite{Kuwano-Jardine}.
Eq~(\ref{1}) is clearly evocative of the Hertz
contact: The energy of compressing two elastic spheres
scales with $(\triangle h)^{2.5}$, where $\triangle h$
is the change in height~\cite{LL7}. Writing the energy
as $\frac12E(\triangle h)^2$, the effective Young
modulus $E\sim (\triangle h)^{0.5}$ vanishes with
$\triangle h$. The physics for the shear modulus is
assumed to be similar.

We postulate Eq~(\ref{1}) to consider its
ramifications -- noting that it should be possible to
derive it employing micro-mechanics~\cite{Goldenberg}:
Although an intricate task, it is not as difficult as
calculating the stress $\sigma_{ij}$ or the compliance
tensor $\lambda_{ijkl}$ directly. Remarkably, assuming
that both moduli vanish with $\delta^{0.5}$, we take
sand to be arbitrarily pliable, not at all
``fragile"~\cite{Cates}. Differentiating the energy
$w$ with respect to $\delta$, $u_s$ yields the
pressure $P$ and shear $\sigma_s$, two scaler
quantities; differentiating it with respect to
$u_{ij}$ yields the complete stress tensor
$\sigma_{ij}$,
\begin{eqnarray}\label{2}
\textstyle P\equiv {\partial w}/{\partial\delta}=\frac
54B\,\delta^{1.5}+\frac 14A\,u_s^2/\delta^{0.5},\\
\sigma_s\equiv{\partial w}/{\partial u_s}=A\,{
\delta}^{0.5}\,u_s.\qquad\quad \label{3}\\ \sigma
_{ij}\equiv {\partial w}/\partial
u_{ij}=-P\delta_{ij}+A\delta^{0.5}u_{ij}^0.\label{3a}
\end{eqnarray}
This is the announced static stress-strain relation.
The first term in $P$ is well-known and considered
characteristic of Hertz contacts. The second term,
accounting both for shear dilatancy and yield, is new.
Dilatancy: Holding $P$ constant, $\delta$ decreases
(and the volume expands) with growing $u_s$. Yield:
For given $u_s$, the compressibility $({\partial
P}/{\partial\delta})^{-1}$ is negative if $\delta$ is
sufficiently small. This implies lack of local
stability, and the system will not remain in the
strained state. It is then, without the capability to
sustain static shear, in a fundamental sense ``fluid"
-- though by no means necessarily Newtonian. In fact,
the energy looses stability even before ${\partial
P}/{\partial\delta}$ turns negative, as the cross
convexity condition $({\partial^2w}/{\partial \delta
^2})({\partial ^2w}/{\partial u_s^2})\geq ({\partial
^2w}/{\partial \delta \,\partial u_s})^2$, or
${u_s^2}/{\delta^2}\le {5B}/{2A}$, also needs to be
met. We saw the significance of instability in a
previous work~\cite{J-L}, but did not realize the
following remarkable point and its consequences:
Rewriting the cross convexity condition by replacing
$\delta, u_s$ with $P,\sigma_s$ leads directly to (the
{\em Drucker-Prager} variant~\cite{sm1} of) the
Coulomb yield condition,
\begin{equation}
|\sigma _s|/P\le \sqrt{4A/5B}. \label{4}
\end{equation}
To account for wet sand, the term $-P_c\delta$ (with
$P_c>0$) is added to the energy $w$. This implies a
force (typically supplied by the water's surface
tension) that compresses the grains even without an
applied pressure. The additional term does not change
the convexity condition, only substitutes $P+P_c$ for
$P$ in Eq~(\ref{2}). As a result, Eq~(\ref{4}) assumes
the {\em Mohr-Coulomb} form, $|\sigma _s|\leq
(P+P_c)\sqrt{4A/5B}$.

As any other elasticity theory, the stress-strain
relation of Eqs~(\ref{2},\ref{3},\ref{3a}) may be
directly solved with appropriate boundary conditions
to obtain a complete stress distribution. Because it
includes yield as given by Eq~(\ref{4}), the {\em
Rankine} states are automatically predicted where
instability is close. And the compliance tensor
$\lambda _{ijkl}$ is obtained by simple
differentiation. Writing $\delta u_{ij}=\lambda
_{ijkl}\delta \sigma _{kl}$ as a vector equation,
$\delta\vec\sigma= \hat M\delta\vec u$, with $\hat M$
a $6\times6$ matrix, we see yield is signified if an
Eigenvalue $m_1$ of $\hat M$ vanishes, with the
Eigenvector $\delta\vec u_1$ indicating the direction
of instability. Explicit calculation shows $\delta\vec
u_1\|(\partial m_1/\partial\vec\sigma)$, implying
$\delta\vec u_1$ is perpendicular to the yield
surface, $m_1(\vec\sigma) \sim|\sigma_s|-
P\sqrt{4A/5B}=0$. If there is no plastic contribution,
this implies flows perpendicular to the yield surface,
a circumstance referred to as the ``associated flow
rule"~\cite{sm1}.

\begin{figure}
\begin{center}
\includegraphics[scale=0.6]{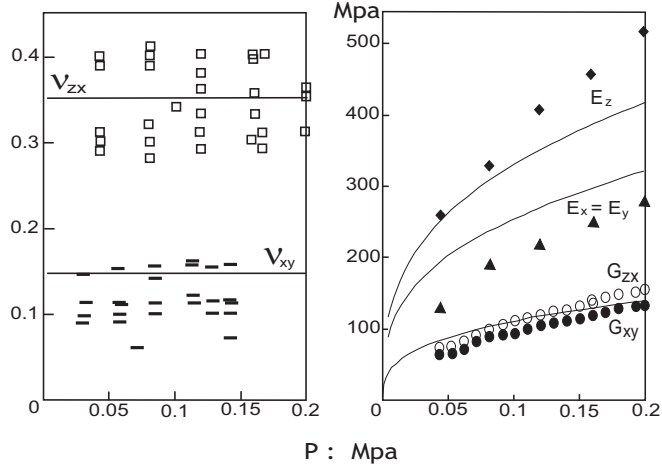}
\end{center}
\caption{The Poisson ratios $\nu_{zx},\nu_{xy}$, the
Young moduli $E_z,E_x=E_y$, and the shear moduli
$G_{zx}, G_{xy}$, measured~\cite {Kuwano-Jardine} with
Ham River sand at $\sigma_{xx}/\sigma _{zz}=0.45$ and
a void ratio of 0.66, compared to the calculated
curves assuming $B=\frac23A=6800$Mpa, with $E_i\equiv
\lambda _{iiii}^{-1}$, $G_{ij}\equiv \frac12\lambda
_{ijij}^{-1}$, $\nu _{ij}\equiv -\lambda
_{iijj}/\lambda_{iiii}$. ($x$, $y$ are horizontal
directions, $z$ the vertical one.) Note these
coefficients are pairwise equal for linear elasticity,
but deviate from each other nonlinearly; theory and
experiment especially agree with respect to the
direction of deviations, ie., the fact that
$\nu_{zx}>\nu_{xy}$, $E_z>E_x=E_y$, $G_{zx}=G_{xy}$.
}\label{nu}
\end{figure}
In view of these results, there can be little doubt
that Eq~(\ref{1}) indeed captures the essence of
granular elasticity. And the remaining question is: To
which extent is it also a quantitative rendition. To
test this, we compare the calculated $\lambda _{ijkl}$
to the data gathered recently~\cite{Kuwano-Jardine},
over a wide range of pressure, shear stress and void
ratio. (Specifying these three variables, the
reversible granular response is unique, showing no
history-dependence.) Fig.~\ref{nu} is a typical plot,
with an overall agreement that further confirms
Eq~(\ref{1}). (The expression for $\lambda _{ijkl}$ is
too cumbersome to be displayed here. It will be given
in a forthcoming single-issue paper containing
extensive comparison.) Note the ratio $A/B$ is fixed
by the Coulomb friction coefficient $\mu_f$, so the
theory has only one overall scale factor, and no
actual adjustable parameter. (The most important
effect missing in Eq~(\ref{1}) is probably
``fabric-anisotropy"~\cite{deGennes}.)

Switching now to a broader context, we proceed to
discriminate between the general feature of the above
theory and those aspects specific to granular
elasticity. This should give us a better appreciation
why Eq~(\ref{1}) is as successful, and also help to
apply the same approach to other jammed systems.
Generally speaking, the energy should be a function of
at least two variables, $u_s$ and $f$, with $f$ being
the one driving the transition, taking place at $f_c$.
In sand, suspensions, and electro-rheological fluids,
$f$ is respectively given by the compression $\delta$,
concentration, and the electric field. Expanding the
energy in $u_s$,
\begin{equation}
\quad w=w_0(f)+\textstyle\frac 12Ku_s^2, \label{5}
\end{equation}
the shear modulus $K$ is a function of $f$, typically
$K\sim(f-f_c)^a$ with $a>0$ in the solid phase
($f>f_c$), and $K\equiv0$ in the liquid one ($f<f_c$).
This dependence is observed in
suspensions~\cite{Trappe}, simulations~\cite{jam} and,
with $a\approx\frac12$, works well for sand. We take
it as an input. Local stability requires $K>0$ and
\begin{equation}
w_0^{\prime \prime }>[(K^{\prime
})^2/K-\textstyle\frac 12K ^{\prime \prime
}]u_s^2\equiv \kappa u_s^2,  \label{6}
\end{equation}
ensuring $w$ is convex in $f, u_s$. Because $\kappa
\sim a(a+1)\times(f_c-f)^{a-2}$ is positive, the
inequality is always violated when $u_s$ becomes
sufficiently large, rendering instability, and hence
the unjamming transition, a generic feature. If $a<2$,
$\kappa$ diverges for $f\to f_c$, and unjamming occurs
at vanishing values of $u_s$ (assuming $w_0^{\prime
\prime }$ remains finite). This ensures the validity
of the expansion of Eq~(\ref{5}).

Considering the jamming transition in the shear-free
limit $u_s\to0$, we identify it -- by analogy to
conventional phase transitions -- as of $n^{\rm th}$
order, if $\partial^i w_0/\partial f^i$ is continuous
for $i<n$, but not for $i=n$. With
$w_0\sim\delta^{2.5}$, sand displays a third-order
jamming transition.

Yield at finite shear, as a result of the energetic
instability, Eq~(\ref{6}), is not an equilibrium
transition, because the liquid phase moves and
dissipates. This may well be compared to raising the
temperature $T$ in a current-carrying superconductor,
such that the metal is pushed into its normal state
carrying a dissipative, ohmic current. In fact, if one
identifies $f$ as $T$, replaces $u_s$ with the
superfluid velocity $v_s$ (and hence $\sigma_s$ with
the current, $j_s=\rho_sv_s$), Eq~(\ref{5}) is valid
for superconductors, and  superfluid helium,
$\frac12Ku_s^2\to\frac12\rho_sv_s^2$, respectively
with $\rho_s\sim T_c-T$ and
$\rho_s\sim(T_c-T)^{2/3}$~\cite{f}. Macroscopically,
jamming and phase transition are clearly hard to tell
apart, and their conceptual difference must be subtle.

\begin{figure}
\begin{center}
\includegraphics[scale=0.4]{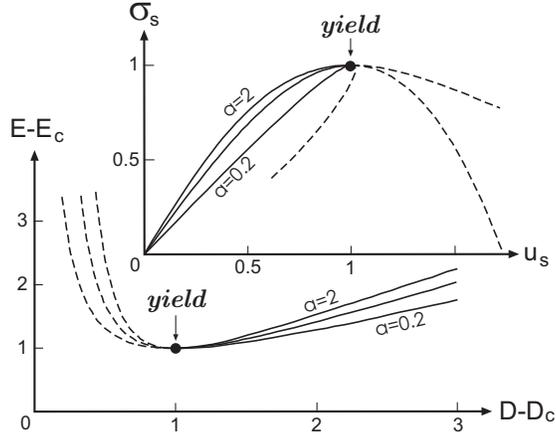}
\end{center}
\caption{Elastic and dielectric properties, including
the yield point, for electro- and magneto-rheological
fluids: Shear stress $\sigma_s$ versus shear strain
$u_s$ at fixed electric field $E$, and $E-E_c$ versus
$D-D_c$ (or $H-H_c$ versus $B-B_c$) at fixed $\sigma
_s$, for the exponents $a=2, 1, 0.2$. Choosing the
dimension of both curves such that the yield points
are at (1,1) render the curves universal -- removing
the dependency on (i) all material parameters other
than the exponent $a$, (ii) $E$ in the upper plot, and
(iii) $\sigma_s$ in the lower one. The relation
$\sigma_s(u_s)$ for $a=1,2$ agree with data from
experiments and
simulations~\cite{Mahon}.}\label{colloid}
\end{figure}
Next, we consider ER and MR (or electro- and
magneto-rheological) fluids, employing them as further
examples for the above notion of jamming. Although
experimental data are as yet not confining enough for
an unambiguous determination of their energy,
plausibility may be drawn on to fill the gap. In ER
fluids, the dielectric displacement $D$ assumes the
role of the transition-driving variable $f$. Writing
the shear-free part of the energy as $w_0=w_1+w_2$, we
take $w_1=\frac 12D^2/\epsilon _1$, accounting for a
linear dielectric relation, and $w_2= -\frac
12\triangle (D-D_c)^2$, assuming that linearity
prevails after the transition at $D_c$. This is a
second-order transition, and the electric field
$E\equiv\partial w/\partial D$ has a kink at $D_c$: We
have $E=D/\epsilon _1$ for $D\leq D_c$, and
$E-E_c=(D-D_c) /\epsilon _2$ for $D>D_c$, with
$1/\epsilon_2= 1/\epsilon _1-\triangle $, $\,E_c\equiv
D_c/\epsilon_1$. (A discontinuity in $E$, or a
first-order transition, was to our knowledge never
reported. Higher order transitions are possible, seem
even likely, but they are not compatible with a linear
constitutive relation after the transition.) $w_2$ is
the condensation energy, so $\triangle$ must be
positive for solidification to take place. (Taking
$D\to B$, $E\to H$ yields the analogous formulas for
MR fluids.) Given $w_0$ and $K=A(D-D_c)^a$, the energy
$w$ of Eq~(\ref{5}) is specified. We calculate the
dielectric relation $E\equiv\partial w/\partial
D|_{u_s}$, elastic relation $\sigma_s\equiv\partial
w/\partial u_s|_{D}$, and rewrite Eq~(\ref{6}) in
terms of $E,\sigma_s$ to obtain the yield condition,
\begin{equation}
|\sigma _s|\le (E-E_c)^{1+\frac a2}
\sqrt{2A\frac{[\epsilon_2 (a+1)]
^{a+1}}{a(a+2)^{a+2}}}. \label{7}
\end{equation}
The exponent $a=1$, or a yield stress $|\sigma _s|\sim
(H-H_c)^{3/2}$ is observed for most
MR-fluids~\cite{Phule-Ginder}. The same value is also
appropriate for a few ER-fluids~\cite{Dvis-Ginder},
though the yield stress is typically
quadratic~\cite{Halsey,Mahon}, $|\sigma_s|\sim
(E-E_c)^{2}$, indicating $a=2$. An ER-fluid capable of
sustaining an unusually high shear strength was
reported~\cite{Wen} to display a nearly linear dependence
of the yield stress, $|\sigma _s|\sim E-E_c$, or $a\ll1$.
For $a=0$, the shear modulus $K$ is independent of the
field, and there is no yield at all. This is the reason
the square root in Eq~(\ref{7}) diverges for $a\to0$, and
possibly explains the observed high yield stress.

Finally, the above approach and results are critically
appraised. (Granular vocabulary is employed for this
purpose, though the statements are equally valid for
ER and MR fluids.) In physics, every microscopic state
has a unique energy. The same holds for macroscopic
ones if we insist on a consistent description. The
macroscopic energy always depends on entropy and
conserved quantities, such as momentum and mass
density. And if the considered system can sustain
static shear stresses, the strain field $u_{ij}$ must
also be included as an independent variable, where
$u_{ij}$ is to be understood, in soil-mechanical
parlance, as the reversible elastic portion of the
strain field.

It is a plain fact that sand piles, if left alone
under gravity, are stable -- in spite of every kind of
infinitesimal perturbations, which are always present.
This demonstrates sand's capability to sustain static
shear and is the reason for including $u_{ij}$.
Irrespective whether a unique displacement field
exists, the elastic description employing $u_{ij}$ is
robust enough to be valid. This is not different from
superfluid helium with vortex lines, in which the
description in terms of the velocity $\boldsymbol
v_s=\frac\hbar m\boldsymbol \nabla\phi$ remains sound,
although the phase field $\phi$ is multivalued.

Given an energy expression $w$, its derivative
$\partial w/\partial u_{ij}$ yields the stress tensor
$\sigma_{ij}$, and its second derivative $\partial^2
w/\partial u_{ij}\partial u_{kl}$ the inverse of the
compliance tensor $\lambda_{ijkl}$. In soil mechanics,
the usual approach consists of postulating the stress
dependence of the 18 independent components of
$\lambda_{ijkl}$ directly, while seeking the account
for the plastic contribution at the same time,
referring to the result as  constitutive
relations~\cite{Kolymbas}. This is quite obviously a
much harder task than finding the one appropriate
scalar expression for the energy $w$ which, even if
heavy-handedly simplified, preserves a large number of
geometric correlation by the mere fact that
$\lambda_{ijkl}$ is obtained via a double
differentiation. We believe this to be the main reason
why the calculated $\lambda_{ijkl}$ stood up so well
when compared to the extensive data
of~\cite{Kuwano-Jardine}.

The expression we proposed in Eq~(\ref{1}) is indeed
the result of weighing simplicity versus accuracy
while stressing the former, and hence is subject to
further scrutiny. As discussed, it includes first of
all an expansion in $u_s$: $w=\frac12 Ku_s^2$ assuming
$K\sim\delta^a$. Starting from $w=\frac
12(B\delta^{2+b}+A{\delta}^a u_s^2)$ in~\cite{J-L}, we
considered the experiments of inclined plane, simple
shear and triaxial test to arrive at $a\approx0.4$,
$b\approx0.5$ giving the best agreement~\cite{a=1}. On
the other hand is the fact that the Coulomb yield
condition, Eq~(\ref{4}), remains unchanged as long as
$a=b$. And it becomes implicit if $a,b$ deviate from
each other -- though the numerical difference is at
first modest. Our tentative choice is $a=b=\frac12$.


\begin{thebibliography}{99}
\bibitem{f} Eq~(\ref{6}) is not valid for superconductors,
because $w(T)$ is a free energy and possesses
different stability conditions.
\bibitem{a=1} Our results cannot be directly
compared to that in~\cite{schall}, because the
pressure contains shear contributions, see
Eq~(\ref{2}). However, if we may neglect them, it is
easy to see that we need to set $a=1$ to obtain a
Young modulus $\sim\sqrt P$.
\end{thebibliography}
\end{document}